\documentclass[twocolumn,           
               showpacs,            
               nopreprintnumbers,     
               aps,                 
               prd,          	    
               letterpaper,             
              groupeaddress,      
               nofootinbib,         
               tightenlines,        
               floats,floatfix,      
               showkeys
               ]{revtex4-1}
               
\usepackage[toc,page]{appendix}
\usepackage{graphicx}
\usepackage{dcolumn}
\usepackage{bm}
\usepackage{amsmath}
\usepackage{amsfonts,amssymb}
\usepackage{soul}
\usepackage{color}
\definecolor{v}{rgb}{0.6, 0.2, 0.8} 
\definecolor{MAGA}{rgb}{0.1, 0.43, 0.75}
\definecolor{jm}{rgb}{0.13, 0.48, 0.64}

\usepackage{xcolor}
\usepackage{enumerate}
\usepackage{float}
\usepackage{subfig}
\usepackage{ulem}

\begin{document}

\title{Einstein-Gauss-Bonnet gravity: is it compatible with modern cosmology?}

\author{Miguel A. Garc\'ia-Aspeitia$^{1,2}$}
\email{aspeitia@fisica.uaz.edu.mx}

\author{A. Hern\'andez-Almada$^3$}
\email{ahalmada@uaq.mx}

\affiliation{$^1$Unidad Acad\'emica de F\'isica, Universidad Aut\'onoma de Zacatecas, Calzada Solidaridad esquina con Paseo a la Bufa S/N C.P. 98060, Zacatecas, M\'exico.}
\affiliation{$^2$Consejo Nacional de Ciencia y Tecnolog\'ia, \\ Av. Insurgentes Sur 1582. Colonia Cr\'edito Constructor, Del. Benito Ju\'arez C.P. 03940, Ciudad de M\'exico, M\'exico}
\affiliation{$^3$Facultad de Ingenier\'ia, Universidad Aut\'onoma de Quer\'etaro, Centro Universitario Cerro de las Campanas, 76010, Santiago de Quer\'etaro, M\'exico.}

\begin{abstract}
Einstein-Gauss-Bonnet (EGB) model is recently restudied in order to analyze new consequences in gravitation, modifying appropriately the Einstein-Hilbert action. The consequences in EGB cosmology are mainly geometric, with higher order values in the Hubble parameter. In this vein this paper is devoted to contribute with extra evidences about its pros and cons of the model from a cosmological point of view. We start constraining the characteristic parameter, $\alpha$, of the EGB model when a cosmological constant as the catalyst for the acceleration is considered. The constrictions are developed at the background cosmology using Observational Hubble Data, Baryon Acoustic Oscillations, Supernovaes of the Ia type, Strong Lensing Systems and the recent compilation of HII Galaxies. Additionally, we implement a statefinder analysis where we found not only a late acceleration but also an early Universe acceleration which is associated with the  parameter $\alpha$. Based on our results of the lensing systems, the Universe evolution never reaches a non accelerated phase, instead it is always presented in an accelerated state, being a possible fault that afflicts the EGB model.
\end{abstract}

\keywords{cosmology, universe acceleration, Einstein-Gauss-Bonnet gravity.}
\pacs{}
\date{\today}
\maketitle

\section{Introduction}

Several theories of gravity have emerged in order to solve the profound conundrums of the Universe evolution like the dark matter (DM) and the dark energy (DE) problems. Centered in the DE problem, the first evidence comes from observations of Type Ia Supernovae (SNIa) \cite{Riess:1998} and confirmed by the acoustic peaks of Cosmic Microwave Background Radiation (CMB) \cite{Planck:2018}, that the Universe is in an accelerated stage and possibly driven by a DE fluid or by a modification to General Theory of Relativity (GR). Among the most accepted candidates are for example the Chaplygin fluids, the cosmological constant, phantom fluids, phenomenological emergent dark energy models, viscous fluids, among other (see the following references \cite{Hernandez-Almada:2018osh,Hernandez-Almada:2020uyr,Hernandez-Almada:2020ulm} for a compilation). On the other hand geometrical extensions to GR are also comprehensive candidates, being the extra dimensional theories \cite{Maartens:2010ar,GarciaAspeitia:2011xv,Garcia-Aspeitia:2016kak,Garcia-Aspeitia:2018fvw}, $f(R)$ theories \cite{Jaime:2013zwa,Jaime:2018ftn}, and unimodular gravity \cite{Perez:2017krv,Perez:2018wlo,Garcia-Aspeitia:2019yni,Garcia-Aspeitia:2019yod}, some of the most important contenders. It is worth notice that in literature, the cosmological constant (CC) which together cold dark matter (DM), baryons and relativistic species, make up the well known standard cosmological model or also called $\Lambda$-Cold Dark Matter model ($\Lambda$CDM), being the CC the preferred component to describe the Universe acceleration. Despite its success, the CC afflicts with severe problems when it is assumed that the CC is caused by quantum vacuum fluctuations, indeed, its theoretical value differs $\sim120$ orders of magnitude \cite{Weinberg,Zeldovich} in comparison with the value obtained by the most precise cosmological observations \cite{Planck:2018}.

Recently, the Einstein-Gauss-Bonnet (EGB) gravity \cite{Glavan:2019inb} has surged, predicting non-trivial contributions to gravitational dynamics and preserving the same number of degrees of freedom for the graviton as GR (other previous studies of EGB can be tracked since the Refs. \cite{Nojiri:2005vv,Amendola_2006}). In cosmology, the EGB framework adds higher order terms in the Hubble parameter to the Friedmann equations which have important consequences mainly in the early epoch of the Universe. Moreover, the continuity equation does not suffer with modification despite the extra terms considered in EGB theory. Matter density perturbations are also studied in EGB context, presenting differences with $\Lambda$CDM in the parameter $\sigma_8$ \cite{Haghani:2020ynl}.
Furthermore, they are tuned with an extra free parameter named $\alpha$ which is expected to be negligible mainly in the late Universe. For instance in \cite{Haghani:2020ynl}, the dimensionless value\footnote{The dimensionless  parameter $\alpha$ can be constructed through $\bar{\alpha}\equiv3\alpha H_0^2$ which will be discussed hereafter.} of $\alpha\to\bar{\alpha}$ is expected to be of the order $10^{-2}$, with the aim to fit with the expected dynamics of Hubble parameter $H$ and the deceleration parameter $q$ (see also \cite{MohseniSadjadi:2020jmc}). Recently, the authors in \cite{Clifton:2020xhc} propose a constriction given by  $0\lesssim\alpha\lesssim10^{8}m^{2}$ based on observations of binary black holes. Other bounds for $\alpha$ from several astrophysical and cosmological phenomena can be checked in \cite{Devi:2020uac,Yang:2020jno,Fernandes:2020rpa,Chakraborty:2020ifg,Banerjee:2020stc}.

Another consequences of EGB gravity are For example, in a robust analysis through dynamical systems applied to cosmology developed in \cite{Chatzarakis:2019fbn} or in studies based on black holes and stellar dynamics in EGB framework, studied in \cite{Devi:2020uac,Yang:2020jno,Fernandes:2020rpa,Chakraborty:2020ifg,Banerjee:2020stc}, which present functional forms of how elucidate the differences among the standard knowledge based in GR and in EGB paradigm. Additionally, studies on strong gravitational systems have been done recently in \cite{Islam:2020xmy,Kumar:2020sag} where they studied several parameters and its correlation with EGB parameter that causes differences with the standard GR for the gravitational lensing by Schwarzschild and charged black holes. Regarding gravitational waves (GW) authors in \cite{Aoki:2020iwm} find a bound for the EGB parameter as $\tilde{\alpha}\lesssim\mathcal{O}(1)$eV$^{-2}$, obtaining one of the first constrictions of $\tilde{\alpha}$ in the EGB model, in addition, Refs. \cite{Clifton:2020xhc} presents therein the bound $\alpha\approx10^{49}$eV$^{-2}$, using the velocity propagation of GW. Nevertheless, it has been found that the model present several mathematical inconsistencies that could rule out the model, being one of them the problem of regularization, which consist in introduce a counter term into the action, being connected with the divergences in the on-shell action (see for instance \cite{Ai:2020peo,Gurses:2020ofy,Lu:2020iav, Fernandes:2020nbq,Mahapatra:2020rds}). Additionally, \cite{Hennigar:2020lsl} show serious faults when the theory is considered to be defined by a set of $D\to4$ solutions of the higher-dimensional Gauss-Bonnet gravity, also notice that in Ref. \cite{Shu:2020cjw} it was found inconsistencies when it is applied to a quantum tunneling process of vacuum. However, we can propound the following questions: Is the EGB model in its current form compatible with modern cosmology? Is it possible to find faults that could rule out the model using cosmological arguments?

In order to respond the previous questions, we gather extra evidence of pros and cons, but now from the cosmological point of view. As far as we know, there is no literature where EGB model (in its current form) has been constrained at cosmological background with the most recent cosmological samples. In this sense, this paper is devoted to revisit the EGB gravity and constrain its main free parameters through the current cosmological observations, like SNIa \cite{Scolnic:2017caz}, Strong Gravitational Lensing (SLS) \cite{Amante:2019xao}, Observational Hubble Data (OHD) \cite{Magana:2017nfs}, Baryon Acoustic Oscillations (BAO) \cite{Nunes:2020hzy}, HII starburst galaxy (HIIG) \cite{Cao:2020jgu}, together with a joint analysis combining the mentioned observations. We compute an appropriate equation of $H^2$, that emerge from the EGB cosmology in order to present the constrictions of the free parameters. Deceleration and jerk parameters are also presented together with a statefinder analysis that discriminate details of EGB in a dynamical context. Our results present an anomalous behavior, not expected in the standard cosmology.

The outline of the paper is as follows. Section \ref{Sec:Framework} is dedicated to present the theoretical framework of the EGB models, focusing in Hubble, deceleration and jerk parameters. Sec. \ref{OC} presents the details of the samples and the methodology to obtain the EGB constraints. Sec. \ref{RES} show the results obtained together with the comparison to the standard $\Lambda$CDM model. Finally in Sec. \ref{CO} we develop the conclusions and outlooks. In what follows we will use units in which $\hbar=c=k=1$, unless we indicate otherwise. 

\section{Theoretical Framework} \label{Sec:Framework}

The action of the EGB gravity can be written in the form \cite{Glavan:2019inb}
\begin{eqnarray}
    &&S_{EGB}[g_{\mu\nu}]=\int d^{d+1}x\sqrt{-g}\Big[\frac{1}{2\kappa^2}(R-2\Lambda)+\mathcal{L}_m\nonumber\\&&+\frac{\alpha}{d-3}\mathcal{G}\Big],
\end{eqnarray}
where $\kappa^2\equiv8\pi G$, $G$ is the Newton constant, $\Lambda$ is an effective cosmological constant, $R$ is the Ricci scalar, $\mathcal{L}_m$  is the matter Lagrangian, $\alpha$ is an appropriate free parameter, $\mathcal{G}=6R^{\mu\nu}_{\;\;\;\;[\mu\nu}R^{\rho\sigma}_{\;\;\;\;\rho\sigma]}$ is the Gauss-Bonnet contribution to the Einstein-Hilbert action and $d+1$ is considered in the limit when $ \lim_{d\to3}d+1$ as presented in \cite{Glavan:2019inb}. Minimizing the action, the field equation can be written as
\begin{eqnarray}
    &&G_{\mu\nu}+\Lambda g_{\mu\nu}+\frac{\alpha}{(d-3)}(4RR_{\mu\nu}-8R_{\mu\alpha}R^{\alpha}_{\nu}\nonumber\\&&-8R_{\mu\alpha\nu\beta}R^{\alpha\beta}+4R_{\mu\alpha\beta\sigma}R^{\alpha\beta\sigma}_{\nu}-g_{\mu\nu}\mathcal{G})=\kappa^2T_{\mu\nu}.
\end{eqnarray}
Notice that when $\alpha=0$, the standard Einstein field equation with a CC is recovered.

In order to study the background cosmology, we assume a flat Friedmann-Lemaitre-Robertson-Walker (FLRW) line element, which
form is $ds^2=-dt^2+a(t)^2(dr^2+r^2d\Omega^2)$ where $d\Omega^2=d\theta^2+\sin^2\theta d\varphi^2$ is the solid angle and $a(t)$ is the scale factor. The energy-momentum tensor is the usual, described by the following tensor equation
\begin{equation}
    T_{\mu\nu}=pg_{\mu\nu}+(\rho+p)u_{\mu}u_{\nu},
\end{equation}
where $p$ and $\rho$ represent the pressure and energy density of the fluid respectively and $u_{\mu}$ is the fluid four-velocity and assumed in a comoving coordinate system. After some manipulations of the previous expressions, the Friedmann equation for EGB reads
\begin{equation}
H^2+3\alpha H^4=\frac{\kappa^2}{3}\sum_i\rho_i+\frac{\Lambda}{3}, \label{Friedmann}
\end{equation}
where we assume a perfect fluid for the energy-momentum tensor and $H\equiv\dot{a}/a$, being the dot a temporal derivative. Moreover, the continuity equation takes its traditional form as
\begin{equation}
\sum_i[\dot{\rho}_i+3H(\rho_i+p_i)]=0.
\end{equation}
In terms of the dimensionless variables, Eq. \eqref{Friedmann} is re-written as
\begin{equation}
    E(z)^2+\bar{\alpha}E(z)^4=\Omega_{m0}(z+1)^3+\Omega_{r0}(z+1)^4+\Omega_{\Lambda 0}, \label{Friedmannnon}
\end{equation}
where $\bar{\alpha}\equiv3\alpha H_0^2$, $\Omega_{i0}\equiv\kappa^2\rho_i/3H_0^2$ and $\Omega_{\Lambda0}\equiv\Lambda/3H_0^2$, being $H_0$ the Hubble parameter at today and it is considered the matter (baryons and dark matter) and relativistic particles (photons and neutrinos), where radiation can be constrained with the expression $\Omega_{r0}=2.469\times10^{-5}h^{-2}(1+0.2271N_{\rm eff})$, where $N_{\rm eff}=3.04$ is the standard number of relativistic particles \cite{Komatsu_2011}. Another important consideration is that $\bar{\alpha}$ is a positive value as inflation demands (see Ref. \cite{Clifton:2020xhc} for details).

In order to constrain the $\bar{\alpha}$ parameter, we divide the problem in two branches through Eq. \eqref{Friedmannnon}. Therefore, if we only consider the branch where we have a real value of $E(z)$, then we have  
\begin{eqnarray}
E(z)^2=\frac{1}{2\bar{\alpha}}\left[\sqrt{1+4\bar{\alpha}\Omega(z)_{std}}-1\right], \label{FriedmannAd}
\end{eqnarray}
where
\begin{equation}
\Omega(z)_{std}\equiv\Omega_{m0}(z+1)^3+\Omega_{r0}(z+1)^4+\Omega_{\Lambda 0}\,,
\end{equation}
is the standard cosmological model. Eq. \eqref{FriedmannAd} is constrained to the condition $E(0)=1$, having the following relation 
\begin{equation}
    \Omega_{\Lambda 0} = \frac{(2\bar{\alpha}+1)^2-1}{4\bar{\alpha}} - \Omega_{m0} - \Omega_{r0}\,. \label{constriction}
\end{equation}
Notice that when $\bar{\alpha}\to 0$, in \eqref{FriedmannAd} the standard Friedmann equation is recovered. In addition, the deceleration can be computed through the $q(z)$ formula, resulting in 
\begin{equation}
    q(z)=\frac{1}{2E(z)^2}\left[\frac{3\Omega_{m0}(z+1)^3+4\Omega_{r0}(z+1)^4}{\sqrt{1+4\bar{\alpha}\Omega(z)_{std}}}\right]-1. \label{q}
\end{equation}
Moreover the jerk parameter can be constructed through the $j\equiv\dddot{a}/aH^3$ function, having
\begin{eqnarray}
&&j(z)=q(z)^2+\frac{(z+1)^2}{2E(z)^2}\frac{d^2E(z)^2}{dz^2}-\frac{(z+1)^2}{4E(z)^4}\nonumber\\&&\times\left(\frac{dE(z)^2}{dz}\right)^2, \label{j}
\end{eqnarray}
where 
\begin{eqnarray}
&&\frac{dE(z)^2}{dz}=\frac{3\Omega_{m0}(z+1)^2+4\Omega_{r0}(z+1)^3}{\sqrt{1+4\bar{\alpha}\Omega(z)_{std}}}, \\
&&\frac{d^2E(z)^2}{dz^2}=-\frac{2\bar{\alpha}[3\Omega_{m0}(z+1)^2+4\Omega_{r0}(z+1)^3]^2}{[1+4\bar{\alpha}\Omega(z)_{std}]^{3/2}}\nonumber\\&&+\frac{6\Omega_{m0}(z+1)+12\Omega_{r0}(z+1)^2}{\sqrt{1+4\bar{\alpha}\Omega(z)_{std}}}.
\end{eqnarray}
Finally, a robust analysis that it is possible to implement via the {\it statefinder analysis}, based in a study in the $\lbrace s,r\rbrace$-plane \cite{Sahni:2002fz,Alam:Stat} is implemented. The main parameters are defined by the following geometric variables
\begin{eqnarray}
    &&r\equiv j=\frac{\dddot{a}}{aH^3}, \\
    &&s=\frac{r-1}{3(q-1/2)},
\end{eqnarray}
where $r\equiv j$, is also the jerk parameter but written in the statenfinder notation. We remark that the $\Lambda$CDM is located at $(s,r)=(0,1)$ in the statefinder space-phase. 

\section{Observational Constraints} \label{OC}

In order to constrain the free parameters we employee the SNIa, BAO, OHD, SLS and HIIG samples (see \cite{Scolnic:2017caz,Amante:2019xao,Magana:2017nfs,Nunes:2020hzy,Cao:2020jgu}). To constrain the EGB parameters, we perform a Markov Chain Monte Carlo (MCMC) analysis, based on the emcee Phyton module \cite{Emcee:2013}, by setting 2500 chains with 250 steps each one. The nburn phase is stopped up to obtain a value of $1.1$ on each free parameter in the Gelman-Rubin criteria \cite{Gelman:1992}. Then, we build a Gaussian log-likelihood as the merit-of-function to minimize
\begin{equation}
-2\log(\mathcal{L}_{\rm data})\varpropto \chi^2_{\rm data},
\end{equation}
for each dataset mentioned previously. Additionally, a joint analysis can be constructed through the sum of them, i.e.,
\begin{equation}
    \chi^2_{\rm Joint}=\chi^2_{\rm SNIa}+\chi^2_{\rm BAO}+\chi^2_{\rm OHD}+\chi^2_{\rm SLS}+\chi^2_{\rm HIIG},
\end{equation}
where subscripts indicate the observational measurements under consideration. The rest of the section is devoted to describe the different cosmological observations.

\subsection{Type Ia Supernovae}

The most recent and largest compilation provided by \cite{Scolnic:2017caz}, contains the observations of the luminosity modulus from 1048 SNIa located in the redshift region $0.01<z<2.3$. The merit of function is constructed as
\begin{equation}
    \chi^2_{\rm SNIa}=(m_{th}-m_{obs})\cdot \rm{Cov}^{-1}\cdot(m_{th}-m_{obs})^T,
\end{equation}
where $m_{th}-m_{obs}$ is the difference between the theoretical and observational bolometric apparent magnitude and $\rm{Cov}^{-1}$ is the inverse of the covariance matrix. The theoretical counterpart is estimated by 
\begin{equation}
    m_{th}=\mathcal{M}+5\log_{10}[d_L(z)/10\, pc],
\end{equation}
where $\mathcal{M}$ is a nuisance parameter and $d_L(z)$ is the dimensionless luminosity distance given by 
\begin{equation}\label{eq:dL}
    d_L(z)=(1+z)c\int_0^z\frac{dz^{\prime}}{H(z^{\prime})},
\end{equation}
where $c$ is the light velocity, recovered in order to maintain the correct units.

\subsection{Baryon Acoustic Oscillations}

BAO are standard rulers, being primordial signatures of the interaction of baryons and photons in a hot plasma on the matter power spectrum in the pre-recombination epoch. Recently, the authors in \cite{nunes:2020} collected 15 transversal BAO scale measurements, obtained from luminous red galaxies located in the region $0.110<z<2.225$. To confront cosmological models to these data, it is useful to build the $\chi^2$-function as
\begin{equation}
\chi^2_{\rm BAO} = \sum_{i=1}^{15} \left( \frac{\theta_{\rm BAO}^i - \theta_{th}(z_i) }{\sigma_{\theta_{\rm BAO}^i}}\right)^2\,,
\end{equation}
where $\theta_{\rm BAO}^i$ is the BAO angular scale and its uncertainty $\sigma_{\theta_{\rm BAO}^i}$ is measured at $z_i$. The theoretical counterpart, $\theta_{th}$, is estimated as
\begin{equation}
    \theta_{th}(z) = \frac{r_{drag}}{(1+z)D_A(z)}\,.
\end{equation}
In the latter, $r_{drag}$ is the sound horizon at baryon drag epoch and $D_A=d_L(z)/(1+z)^2$ is the angular diameter distance at $z$, with $d_L(z)$ defined in (\ref{eq:dL}). Finally, we use the $r_{drag}=147.21 \pm 0.23$ reported by \cite{Planck:2018}.

\subsection{Observational Hubble Data}

The Observational Hubble Data (OHD) are a cosmological model independent measurements of the Hubble parameter $H(z)$. We consider the OHD compilation provided by \cite{Magana:2017nfs} which contains 51 points given by the differential age (DA) tool and BAO measurements, within the redshift region $0<z<2.36$. Hence, the chi square function for OHD can be written as
\begin{equation}
    \chi^2_{{\rm OHD}}=\sum_i^{51}\left(\frac{H_{th}(z_i)-H_{obs}}{\sigma^i_{obs}}\right)^2,
\end{equation}
where $H_{th}(z)$ and $H_{obs}(z_i)\pm\sigma_{obs}^i$, are the theoretical and observational Hubble parameter at the redshift $z_i$.

\subsection{Strong Lensing Systems}

The Einstein radius of a lens described by the Singular Isothermal Sphere (SIS), is defined by 
\begin{equation}
    \theta_E=4\pi\frac{\sigma_{SIS}^2D_{ls}}{c^2D_s},
\end{equation}
where $\sigma_{SIS}$ is the velocity dispersion of the lens galaxy, $D_s$ is the angular diameter distance to the source, and $D_{ls}$ is the angular diameter distance from the lens to the source. While the first one is obtained by $D_s=d_L(z)/(1+z)^2$, the latter is estimated by
\begin{equation}
         D_{ls}(z)=\frac{c}{1+z}\int_{z_l}^{z_s}\frac{dz^{\prime}}{H(z^{\prime})},
\end{equation}
where $z_l$ ($z_s$) is the redshift of the lens (source). Hence, it is possible to define a theoretical distance ratio $D^{th}\equiv D_{ls}/D_s$ and the observable counterpart as $D^{obs}=c^2\theta_E/4\pi\sigma^2$. Therefore, the merit of function for SLS is
\begin{equation}
    \chi^2_{\rm SLS}=\sum_i^{205}\frac{[D^{th}(z_l,z_s)-D^{obs}(\theta_E,\sigma^2)]^2}{(\delta D^{obs})^2},
\end{equation}
where 
\begin{equation}
    \delta D^{obs}=D^{obs}\left[\left(\frac{\delta\theta_E}{\theta_E}\right)^2+4\left(\frac{\delta\sigma}{\sigma}\right)^2\right]^{1/2},
\end{equation}
being $\delta\theta_E$ and $\delta\sigma$ the uncertainties of the Einstein radius and velocity dispersion respectively.

\subsection{HII Galaxies}

As it is discussed in \cite{Cao:2020jgu}, the HIIG can be used as an alternative form to constrain cosmological models due to the correlation between the observed luminosity and the velocity dispersion of the ionized gas. This mentioned correlation can be written as
\begin{equation}
    \log L=\beta\log\sigma+\gamma,
\end{equation}
where $L$ is the luminosity, $\sigma$ the velocity dispersion, $\gamma$ and $\beta$ are the intercept and slope functions respectively \cite{Cao:2020jgu}. Therefore, the distance modulus takes the form
\begin{equation}
    \mu_{obs}=2.5\log L-2.5\log f-100.2,
\end{equation}
where $f$ is the flux emitted by the HIIG. On the other hand, the theoretical distance modulus is
\begin{equation}
    \mu_{th}(z)=5\log d_L(z)+\mu_0,
\end{equation}
being $d_L(z)$ the luminosity distance defined in (\ref{eq:dL}) and $\mu_0$ a nuisance parameter.
Hence, motivated by the recent collection provided by \cite{Cao:2020jgu} containing a total of 153 HIIG measurements, we constrain the parameter phase-space ${\bf \Theta}=(h, \Omega_{m0}, \bar{\alpha})$ of EGB model by building the following figure-of-merit
\begin{equation}\label{eq:chiHIIG}
    \chi^2_{{\rm HIIG}} = A - B^2/C\,,
\end{equation}
where
\begin{eqnarray}
    A &=& \sum_{i=1}^{153} \left( \frac{\mu_{th}(z_i)-\mu_{obs}^i}{\epsilon_{\mu_{obs}^i}}\right)^2 \,,\\
    B &=& \sum_{i=1}^{153} \frac{\mu_{th}(z_i)-\mu_{obs}^i}{\epsilon_{\mu_{obs}^i}} \,, \\
    C &=& \sum_{i=1}^{153} \frac{1}{(\epsilon_{\mu_{obs}^i})^2}\,.
\end{eqnarray}
being $\epsilon_i$ is the uncertainty of the $i_{th}$ measurement.

\section{Results} \label{RES}

\begin{figure}
\includegraphics[width=0.485\textwidth]{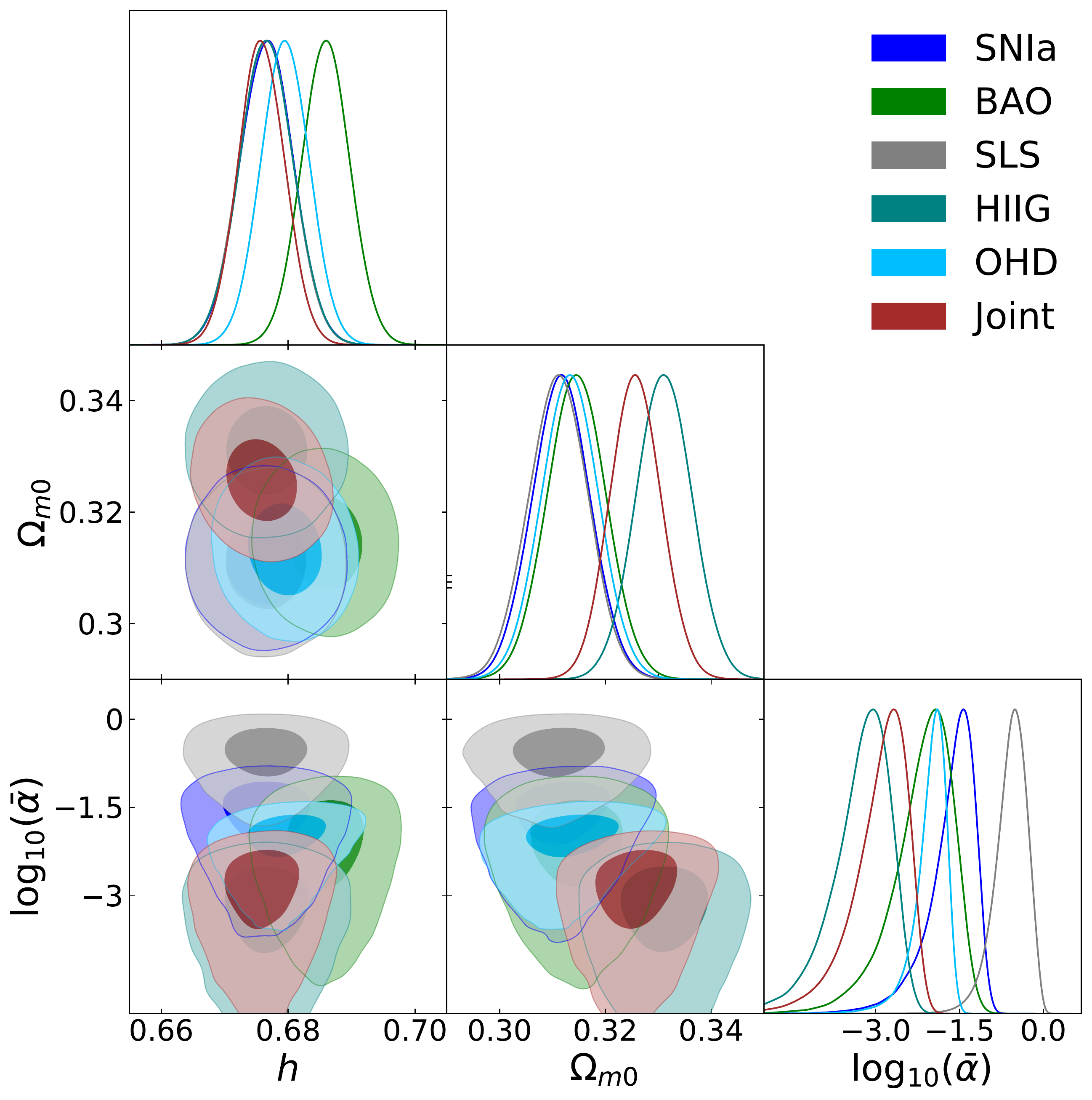}
\caption{1D posteriors distributions and 2D contours of the free parameters for EGB model at $1\sigma$ and $3\sigma$ CL (from darker to lighter respectively).}
\label{fig:Contours}
\end{figure}

\begin{figure}
\includegraphics[width=0.485\textwidth]{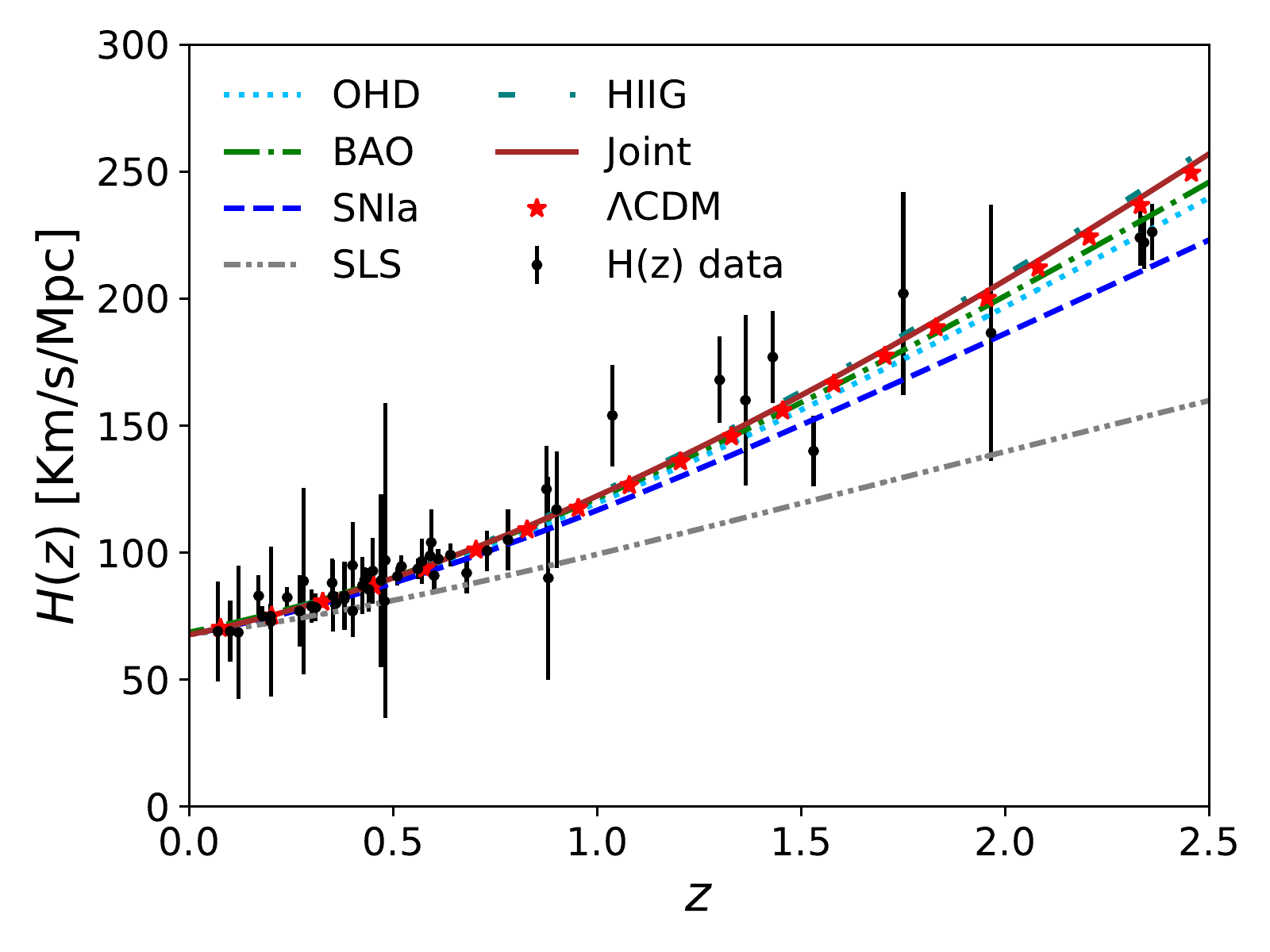}
\caption{Best fits for each data sample and joint, together with the comparison with the standard $\Lambda$CDM model.}
\label{fig:H}
\end{figure}

\begin{figure}
\includegraphics[width=0.485\textwidth]{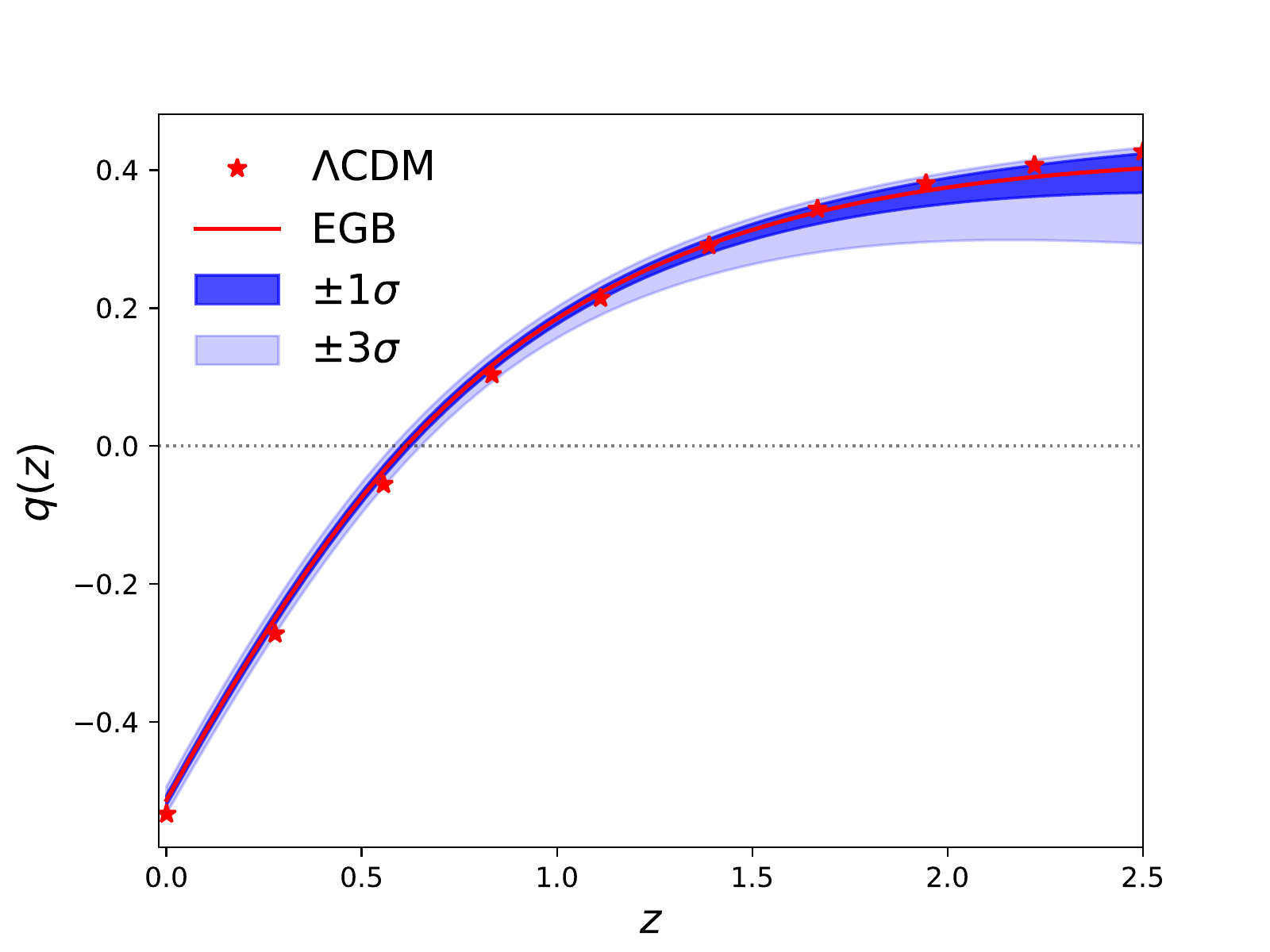} \\
\includegraphics[width=0.455\textwidth]{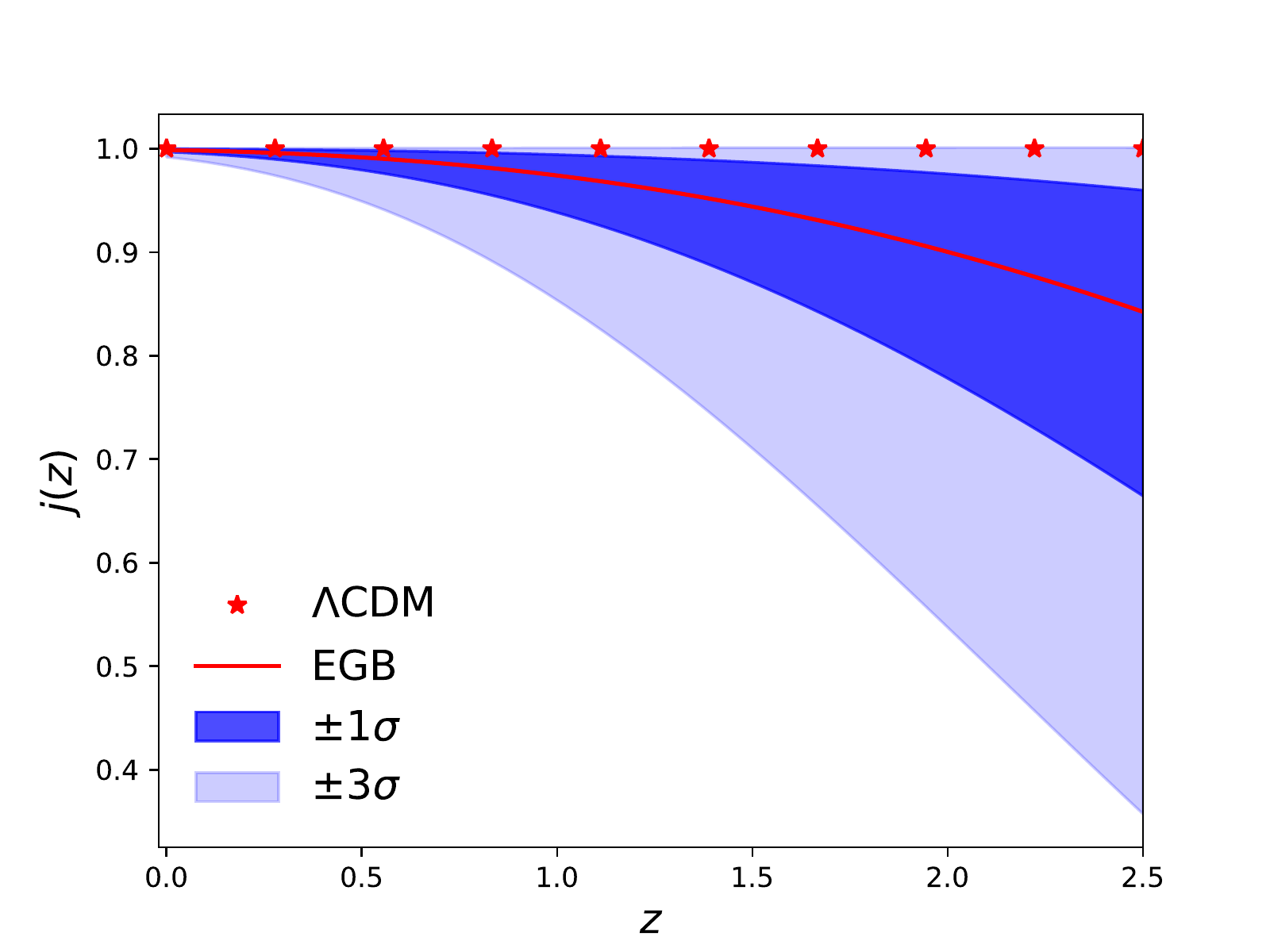}
\caption{Reconstruction of deceleration and jerk parameters (red solid line) using the joint analysis. Inner (outer) bands represent the uncertainties at $1\sigma$ ($3\sigma$) CL. Magenta star markers represent $\Lambda$CDM model.}
\label{fig:qj}
\end{figure}

Our cosmological constrictions applied to EGB model are presented in Fig. \ref{fig:Contours} which the 2D regions correspond to 68\%  ($1\sigma$) and 99.7\% ($1\sigma$) confidence level (CL)  for darker and lighter contours respectively for each data sample.  Furthermore, the central values of the free model parameters with their uncertainties at $1\sigma$ are summarized in Table \ref{tab:bestfits} according to each data and the combined data (joint analysis). The joint analysis give us a value of $\bar{\alpha}$ of the order $\sim10^{-3}$, which is the most restricted bound on $\bar{\alpha}$ through cosmological observables and competitive with those obtained by \cite{Haghani:2020ynl}. These confidence contours also remark that there is not a tension with cosmological data at least at background level, but it is needed a CC to obtain the expected Universe dynamics, mainly at low redshift. However the mystery of the origin of the CC is still an open question. Moreover, it is presented in Fig. \ref{fig:H} the best fits for the evolution of the Hubble parameter in terms of the redshift for the four cosmological samples and joint; a comparison with the standard cosmological model it is shown, together with the $H(z)$ data. In addition, we show in Figs. \ref{fig:qj} the deceleration and jerk parameters and their comparison with $\Lambda$CDM. As it is interesting to observe from the jerk reconstruction, the EGB model has as upper limit the $\Lambda$CDM in the past ($z>0$), having a convergence to this model at $z=0$. Therefore a possible conclusion is due to the correction terms to $\Lambda$CDM caused by EGB gravity, CC mimics dynamical characteristics instead to have $j^{\Lambda CDM}=1$ for times $z<2.5$. Additionally, we estimate a yield value of the deceleration-acceleration transition at $z_t = 0.612^{+0.012}_{-0.012}$. At current epochs, we obtain $q_0=-0.513^{+0.007}_{-0.007}$ and
$j_0=0.999^{+0.001}_{-0.002}$ for the deceleration and jerk parameters, respectively. In case of $\Lambda$CDM, the transition redshift is estimated at $z_t^{\Lambda CDM}=0.642^{+0.014}_{-0.014}$, which corresponds a deviation of $0.92\sigma$, while the deceleration parameter today is $q_0^{\Lambda CDM}=-0.63^{+0.02}_{-0.02}$ for $\Lambda$CDM obtaining a deviation of $3.11\sigma$. Additionally, another useful comparison is with cosmological viscous models like those studied in \cite{Herrera-Zamorano:2020rdh}, where $q_0^{CVM}=-0.568^{+0.018}_{-0.021}$, $j_0^{CVM}=1.058^{+0.039}_{-0.033}$,  $q_0^{PVM}=-0.472^{+0.064}_{-0.056}$ and  $j_0^{PVM}=0.444^{+0.344}_{-0.394}$ for the constant (CVM) and polynomial viscous models (PVM) respectively. In comparison, we obtain a deviations up to $1.61\sigma$ ($0.91\sigma$) between CVM (PVM) and EGB model.

\begin{figure}
\includegraphics[width=0.485\textwidth]{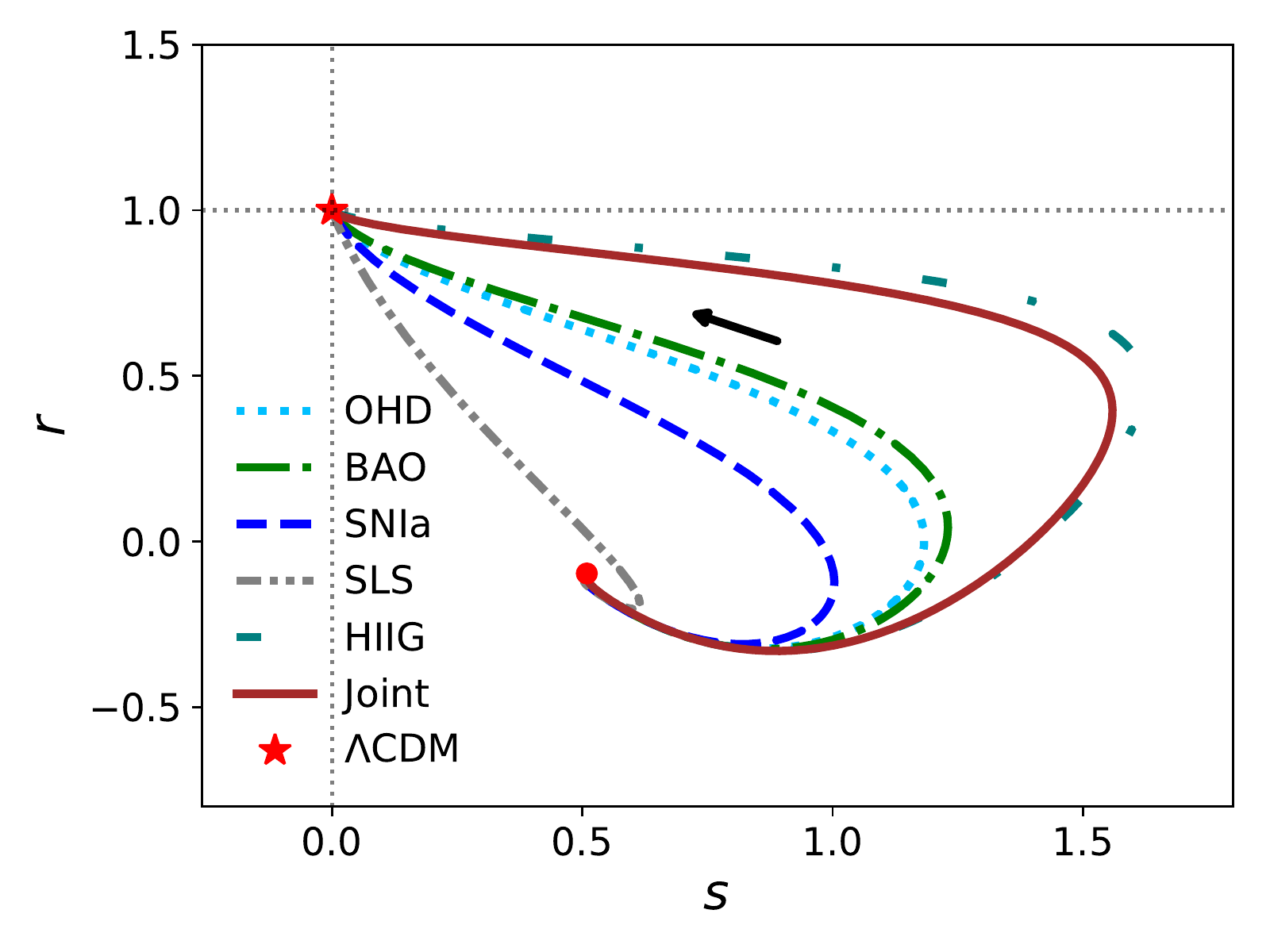} \\
\includegraphics[width=0.455\textwidth]{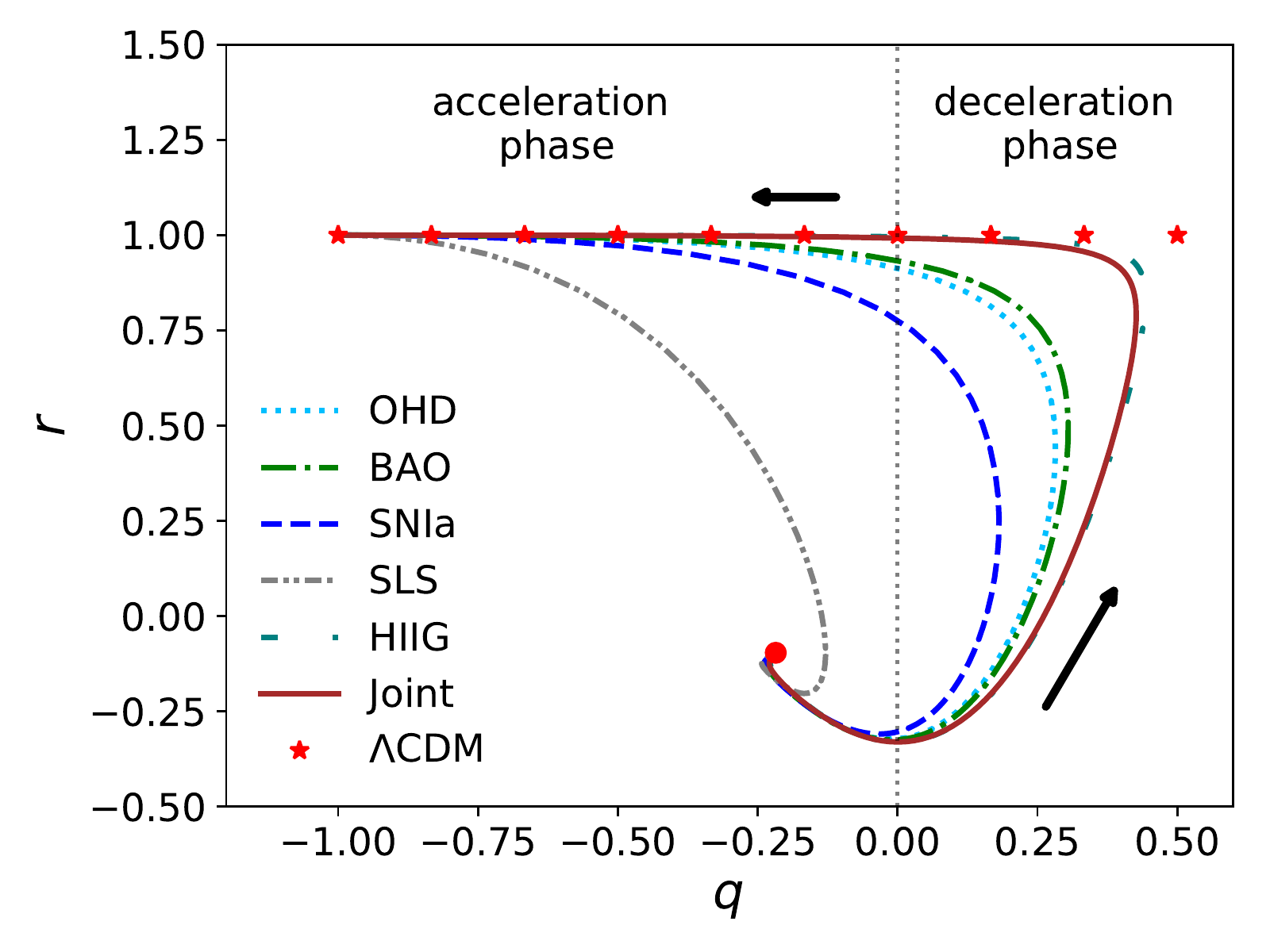}
\caption{Top panel: Evolution of EGB model in the statefinder phase-space. At the future, EGB model converges to $\Lambda$CDM. Bottom panel: jerk {\it vs} deceleration parameter, showing a division between an accelerated and decelerated Universe. Both panels correspond to the redshift range $-1<z<1000$.}
\label{fig:SF}
\end{figure}

Regarding statefinder analysis, we show in Figs. \ref{fig:SF} the phase-state for r-s and r-q respectively. Notice the convergence to the $\Lambda$CDM attractor point for lower values of $\bar{\alpha}$, as it is expected. Furthermore, it is important to remark that this model predicts an early accelerated phase --not only the accelerated phase at $z\sim0.6$-- (see Figs. \ref{fig:SF}), which could generate problems for the established knowledge of cosmology, in particular, for important phases of the Universe like structure formation, nucleosynthesis, reionization, among others. Based on our joint analysis, the earlier acceleration-deceleration transition happens around $z_t\approx17.2$ which coincides with reionization epoch. Notice also that for values of $\bar{\alpha}\sim10^{-9}$,  this transition is approximately moved to the photon decoupling era ($z\approx1100$). Moreover, lower values for $\bar{\alpha}$ imply a no so long early accelerated phase, which could have important consequences, notice however that through cosmological constrictions always exist this early acceleration (see Figs \ref{fig:SF})\footnote{Notice that when $\bar{\alpha}\to0$ the early acceleration tends to infinity.}. Other complementary constrictions developed by astrophysical events could reduce the presence of $\bar{\alpha}$ avoiding the early Universe accelerations predicted by our cosmological constrictions. 

\begin{table*}
\caption{Best fitting values of the free parameters for the EGB model with the different samples used in this paper.}
\centering
\begin{tabular}{|cccccc|}
\hline
Sample     &    $\chi^2$     &  $h$ & $\Omega_{m0}$ & $\bar{\alpha}$ &  $\mathcal{M}$          \\
\hline
 OHD   & $25.8$ & $0.677^{+0.004}_{-0.004}$ & $0.312^{+0.005}_{-0.005}$ & $0.011^{+0.007}_{-0.005}$ & --  \\ [0.7ex]
 BAO    & $40.7$ & $0.686^{+0.004}_{-0.004}$ & $0.315^{+0.006}_{-0.006}$ & $0.008^{+0.014}_{-0.006}$ & -- \\ [0.7ex]
SNIa    & $39.8$ & $0.677^{+0.004}_{-0.004}$ & $0.312^{+0.005}_{-0.005}$ & $0.028^{+0.028}_{-0.018}$ &  $-19.400^{+0.016}_{- 0.016}$   \\ [0.7ex]
SLS     & $577.8$  & $0.676^{+0.004}_{-0.004}$ & $0.311^{+0.006}_{-0.006}$ & $0.281^{+0.218}_{-0.143}$  & --    \\ [0.7ex]
HIIG    & $2269.5$ & $0.677^{+0.004}_{-0.004}$ & $0.331^{+0.005}_{-0.005}$ & $0.0006^{+0.0011}_{-0.0005}$ & --    \\ [0.7ex]
 Joint  & $6181.9$ & $0.676^{+0.004}_{-0.004}$ & $0.326^{+0.005}_{-0.005}$ & $0.001^{+0.002}_{-0.001}$ & $-19.400^{+0.012}_{- 0.012}$ \\ [0.7ex]
\hline
\end{tabular}
\label{tab:bestfits}
\end{table*}

\section{Conclusions and Outlooks} \label{CO}

In this paper, we present the constraints for EGB free parameters through the most recent cosmological observation. Our results from the Joint analysis, show that the EGB free parameter $\bar{\alpha}$, must be of the order $0.001^{+0.002}_{-0.001}$, therefore we have that $\alpha\approx10^{-3}/3H_0^2=(1.604^{+0.017}_{-0.018})\times10^{62}$eV$^{-2}$, which is the best constriction up to now using cosmological data at the background level. Our results from cosmological point of view complement also to those collected in \cite{Clifton:2020xhc}, where they found the bound $0\lesssim\alpha\lesssim2.57\times10^{21}$eV$^{-2}$, from binary black holes systems and $\alpha\approx10^{49}$eV$^{-2}$, from velocity propagation of gravitational waves \cite{Clifton:2020xhc,TheLIGOScientific:2017qsa}.

In addition, the different cosmological observations are not in tension with the cosmological EGB model, but it is necessary to include a CC to drive the late accelerated expansion phase of the Universe. When the deceleration and jerk parameters were analyzed, we found there is a deceleration-acceleration transition at $z_t = 0.612^{+0.012}_{-0.012}$ which is in  agreement with the standard cosmological model. On the other hand, the analysis of the jerk parameter show us that the CC can mimic a dynamical evolution under the EGB formalism, without adding complexity in the EoS equation associated to DE.  

Moreover, the statefinder analysis remarks an accelerated Universe phase in early epochs, which could have affectations in nucleosynthesis or reinoization epochs. In particular for the joint analysis results, we have a transition to an accelerated Universe at $z_t\approx17.2$, which is into the reionization epoch. As it is possible to observe, large values of $\bar{\alpha}$ generate that the early universe acceleration stay long time, even, never disappears from the scenario evolution like in the case of SLS constrictions. We observe that for values $\bar{\alpha}\sim10^{-9}$, the accelerated phase happens approximately at photon decoupling epoch, possible modifying its physics. On the other hand, such earlier acceleration phased may play an important role to explain the inflationary cosmology that could be explored in future works. In this sense, further studies through astrophysical events, such as black holes or neutron stars, are needed to obtain strong constrictions of the parameter $\bar{\alpha}$. 

Regarding the regularized version of the models presented in Refs. \cite{Lu:2020iav,Feng:2020duo,Hennigar:2020lsl,Aoki:2020iwm}, this is an alternative approach to alleviate the mentioned problems and it is possible to apply the same methodology, constraining the free parameters. As a complement, a statefinder analysis is necessary in order to study if the regularized version has the same afflictions associated with the unexpected acceleration in the early epochs of the universe.

Finally, summarizing the previously mentioned afflictions of the model with our findings about an early acceleration that should affect some important process, could be ruled out the model in its present form. We suggest, that the EGB model in its current form, must be mathematically restructured and with this, it is possible to develop again a background cosmological analysis (like the presented in this paper) and extend it to the perturbative level.

\begin{acknowledgements}
We thank the anonymous referee for thoughtful remarks and suggestions. M.A.G.-A. acknowledges support from SNI-M\'exico, CONACyT research fellow, CONICYT REDES (190147), COZCyT and Instituto Avanzado de Cosmolog\'ia (IAC). A.H.A. thanks to the SNI-M\'exico, CONACyT for support. \\
\end{acknowledgements}

\bibliography{librero1}

\end{document}